# Shielding electrostatic fields in polar semiconductor nanostructures

G. M. O. Hönig*, S. Westerkamp, A. Hoffmann, and G. Callsen*

*Institut für Festkörperphysik, Technische Universität Berlin, Hardenbergstr. 36, 10623 Berlin, Germany*

*gerald.hoenig@physik.tu-berlin.de, callsen@tu-berlin.de

**Modern opto-electronic devices are based on semiconductor heterostructures employing the process of electron-hole pair annihilation. In particular polar materials enable a variety of classic and even quantum light sources, whose on-going optimisation endeavours challenge generations of researchers. However, the key challenge - the inherent electric crystal polarisation of such materials - remains unsolved and deteriorates the electron-hole pair annihilation rate. Here, our approach introduces a sequence of reverse interfaces to compensate these polarisation effects, while the polar, natural crystal growth direction is maintained provoking a boost in device performance. Former research approaches like growth on less-polar crystal planes or even the stabilization of unnatural phases never reached industrial maturity. In contrast, our solution allows the adaptation of all established industrial processes, while the polarisation becomes adjustable; even across zero. Hence, our approach marks the onset of an entire class of ultra-fast and efficient devices based on any polar material.**

Semiconductor structures with thicknesses of a few nanometres are commonly used to confine electrons and holes in close vicinity, which increases the electron-hole pair annihilation rate governing the photon generation. As a result, any dense semiconductor band structure is transformed into distinct electronic levels for electrons and holes, whose spacing can be simply tailored by the extent of the nanostructure.[1] Hence, not only the semiconductor material for itself determines the colour of the emitted photons based on its individual band gap, also any structural size variation alters the optical properties.

For instance, nowadays light emitting diodes (LEDs) and laser diodes (LDs) for visible and ultraviolet light emission are predominantly based on group-III nitride heterostructures[2] exhibiting a wurtzite crystal structure. All the advantages of this material system like an astonishing robustness, brilliance, and integrability come at the cost of a piezo- and pyroelectric polarisation along the natural [0001] growth direction.[3] Generally, any stress-induced deformation of such polar crystals yields macroscopic charges at the crystal surfaces - a well known phenomenon for everyday devices like piezo ignition lighters. While the pyro-electricity is caused by the particular symmetry of the crystal, the piezoelectricity originates from strain evoked by hetero-interfaces of such materials with different lattice constants, altogether creating huge electrostatic fields (MV/cm) across the optically active regions.[4] Hence, the beneficial confinement effect for electrons and holes is counterbalanced by a field-induced spatial separation, which spoils their pivotal overlap and ultimately the annihilation rate governing the light generation.[5]

Here, we demonstrate that our Internal Field Guarded Active Region Design (IFGARD)[6] is able to lock the polarisation fields out of any active region, while the polar and industrial most relevant growth direction is maintained, yielding a boost in electron-hole annihilation rates and therefore device efficiency by orders of magnitude. Previous attempts to tackle polarisation fields in affected materials did not reach industrial-relevance. Stabilizing the cubic phase of nitrides[7] might be an approach of scientific elegance, however, the crystal quality deteriorates,[8,9] turning the bright nitrides into rather dim emitters.[10] Similarly, the realisation of numerous alternative growth directions of nitride materials can (partially) avoid the polarisation fields, but again suffers from a strongly reduced crystal quality besides additional technological challenges.[11] In contrast, the IFGARD can bear on decades of research to improve the quality of strongly polar materials based on their most natural growth direction. The efforts to suppress any radiative as well as non-radiative losses has nowadays already led to record quantum efficiencies;[12,13] a tedious and costly progression for polar material forming the basis for the IFGARD. Hence, our approach will combine the advantages of two worlds - well-established, high quality, polar material *and* high electron-hole pair annihilation rates.

In this article, the focus rests on two technologically most relevant types of nano heterostructures: two-dimensional quantum wells (QWs) and zero-dimensional quantum dots (QDs). Generally, the IFGARD does not only serve classical applications as LEDs and LDs, but will even break new ground for efficient and ultra-fast quantum light sources based on individual QDs.

## Field guarding in quantum dots

The IFGARD is based on a simple, but at the same time counterintuitive idea. Commonly, the emissive, active material region (e.g. QD, QW) of devices features a smaller band gap than the matrix material in its surrounding providing the beneficial carrier confinement effect. In this situation, as exemplified in **Fig. 1 a**, any conventional device design strives to avoid additional layers that solely comprise the material of the active region to avoid light reabsorption. Hence, it appears as a ludicrous design to encapsulate this sandwich by the active region material as shown in **Fig. 1 d**. In this counterintuitive design a significant fraction of the emitted light gets reabsorbed by the, so-called, guard layers, casting doubt on the usability of the device - at first sight. However, guard material thicknesses below the emitted wavelength only absorb a well tolerable amount of emitted photons as discussed in the Supplementary Information (**SI**) - an effect that is by far overcompensated by the advantages of the IFGARD aiming to annihilate the detrimental, polarisation-induced, electric fields based on its particular design.

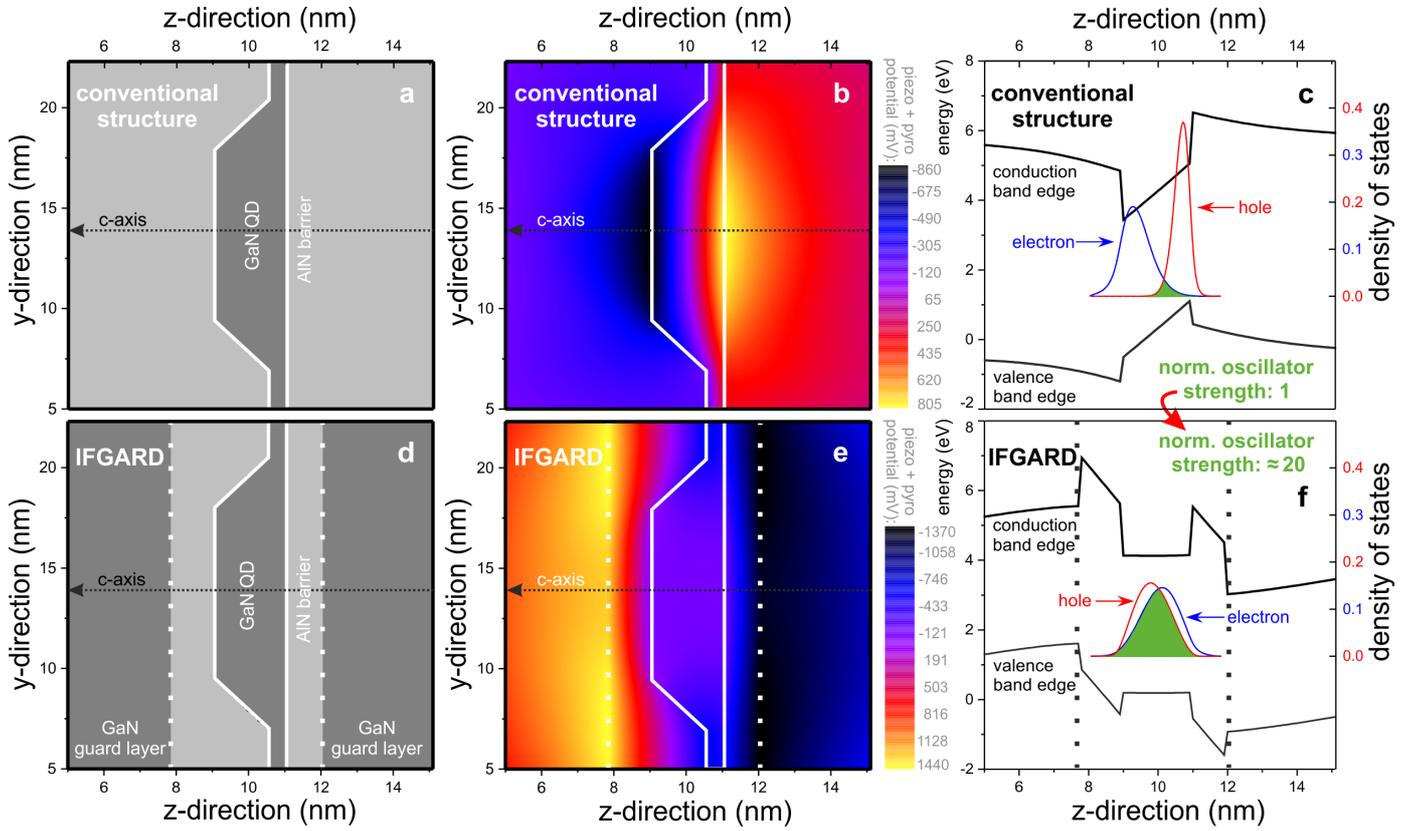

**Figure 1 | IFGARD QD.** Comparison between a conventional quantum dot (QD) structure (1st row) and a QD comprising the Internal Field Guarded Active Region Design (IFGARD, 2nd row). From left to right: the particular layer sequence is exemplified for the GaN/AlN case as illustrated in the corresponding 2D-scans in (a) and (d). The contour-plots in (b) and (e) show the sum of the piezo- and pyroelectric potential for the conventional and the IFGARD QD structure. As a consequence of such particular potential distributions, different conduction and valence band edge profiles are obtained for a linear scan through the QD centre along the c-axis as depicted in (c) and (f). While the conventional QD structure exhibits a prominent potential gradient (yellow → black) inside of the QD (b), the IFGARD QD features a constant potential inside of the QD as evidenced by the purple colouring in (e). As a result, flat-band conditions are achieved inside of the IFGARD QD in contrast to a strong band-edge tilt for the conventional case (f, c). Consequently, the potential gradient inside of the conventional QD structure separates the charge carriers as shown by the electron (blue) and hole (red) density of states in (c). In contrast, a drastically increased electron-hole overlap is obtained for the IFGARD QD (f) causing a beneficial boost in electron-hole oscillator strength and recombination rate.

In order to exemplify the IFGARD, we first choose a GaN QD embedded in AlN - a selection that does not restrict the general applicability of the entire concept to a specific material system and/or nanostructure. **Figure 1** summarizes the major differences between a conventional GaN QD and its IFGARD counterpart in the first and second row, focussing from left to right on the composition, the polarisation fields, and the band structure. Here, the horizontal c-axis denotes the most favourable, natural [0001] growth direction of III-nitrides. **Fig. 1 a** shows a GaN QD with a height along this c-axis of 2 nm (dark grey) embedded in a matrix of AlN (light grey), while the IFGARD equivalent features thin AlN barriers and additional GaN guard layers as depicted in **Fig. 1 d**. A significant interface charge built-up occurs at the AlN/GaN/AlN interfaces, yielding a huge polarisation gradient with a potential drop of ≈ 1.7 V for the conventional case - right across the QD as shown in the colour-coded image of **Fig. 1 b**. Naturally, the associated polarisation potential overlays the band structure, provoking the band edges to be tilted right along the horizontal c-axis as shown in **Fig. 1 c**. As a result, not only a red-shift of the emission wavelength occurs, but also the electron and hole are spatially separated along the c-axis, lowering their overlap and subsequently the electron-hole pair annihilation rate. **Figure 1 c** illustrates this matter based on the electron (blue) and hole (red) density of states (profiles along the c-axis through the QD centre) and the corresponding overlap (coloured in green). The entire phenomenon that counteracts the confinement-induced blue-shift of the QD emission is known as the Quantum-Confined Stark Effect (QCSE) and has been studied in great profusion in the last decades.[14-17] It is exactly this QCSE that researchers sought to overcome by e.g. stabilizing the cubic crystal structure phase or by realizing numerous alternative growth directions of III-nitrides.

Generally, the same charge built-up occurs for the IFGARD case depicted in **Fig. 1 d**. However, due to the inclusion of the guard layers, the polarisation potential gradient is now suspended from the QD. By adding two additional GaN/AlN interfaces as described by the IFGARD, one can suppress the electric field inside of the QD as depicted in **Fig. 1 e**. Here, the constant purple colouring of a major fraction of the QD approves constancy for the sum of the piezo- and pyroelectric polarisation potential - the main benefit of the IFGARD. As a result, one obtains flat conduction and valence band edges within the QD, a strongly reduced charge carrier separation along the c-axis, and, as a direct consequence, an enhanced electron-hole overlap as shown in **Fig. 1 f**. Therefore, in comparison to the conventional case, the IFGARD raises the directly related oscillator strength by a factor of 20 for the common QD dimensions assumed in **Fig. 1**. This improvement directly translates to a factor of 20 in the rate of emitted, single photons from such a GaN QD. Please note that all detailed information regarding the simulations (8-band-k·p implementation for wurtzite materials like nitrides) can be found in the **SI**, cf. **Fig. S1**. Here, also the particle interactions are considered for the aforementioned electron-hole pairs (Hartree-Fock treatment) in order to approximate the corresponding two-particle state known as an exciton.

### Tailoring the internal field

The presented drastic changes regarding the emission characteristics are predominantly caused by the polarisation effects inherent to the crystal lattice and not by the influence of strain on the band structure - in this context an almost negligible,[19-21] but nevertheless still considered effect (see the **SI** for details). Therefore, we focus on the electric potential instead of the band edge profiles in order to illustrate the IFGARD effect in the following. Annihilating the QCSE by cancelling out the electric fields generated by the interface charges positioned on the opposite sides of the QD and barriers always exhibits the most tremendous effect. However, some fine tuning of the AlN barrier thicknesses is needed in order to reach a fully optimised field cancellation for QDs not only due to their top and bottom facets of different size but also due to their inclined side facets. Please note that these *top* and *bottom* facets correspond to the *left* and *right* GaN/AlN interface of the QD in **Fig. 1 and 2** in order to allow a convenient comparison.

**Figure 2** focuses on the influence of structural IFGARD parameters on the polarisation potential within another, here, 3-nm-high (h) QD shown in **Fig. 2a**. By varying the top AlN barrier thickness (t, red) above and the bottom barrier thickness (b, blue) below the QD, the gradient of the built-in electric potential drastically varies as plotted in **Fig. 2b**. By symmetrically decreasing both barrier widths (t = b), the potential gradient evolves from a drop (blue curve) within the QD for thick barriers (regarding a positive probe charge) to a corresponding rise for thinner barriers (red curve). AlN barrier thicknesses in between 1.5 nm and 2.0 nm (t = b) yield the smallest slopes for the potential trends as long as symmetric barriers are considered. However, the potential drop inside of the QD can be reduced even further if different barrier thicknesses are considered (t ≠ b). We find the combination of 1.5-nm- and 2.0-nm-thick AlN barriers to be ideal for reducing the absolute potential drop from the top to the bottom edge of the 3-nm-high IFGARD QD down to 5 mV. Interestingly, the inversion of the stack sequence (t ↔ b) does neither significantly alter the gradient, nor the particular trend for the electric potential, as evidenced by the black and green curves in **Fig. 2 b**. Here, only the bottom barrier thickness regulates the absolute value of the potential inside of the QD in regard to an arbitrarily chosen zero. As soon as the flat-potential conditions are approached, a potential bowing becomes apparent originating from the piezoelectric polarisation, which is caused by the particular strain distribution inside of the QD. It is of utmost importance to note that exactly the same 3-nm-high QD embedded in a conventional structure is affected by a total potential drop of -2112 mV as indicated in **Fig. 2 b** (dashed, grey line) in contrast to the optimum of -5 mV. Therefore, we use exactly this straightforwardly accessible total potential drop (PD) as a convenient measure for the degree of internal field guarding due to the IFGARD.

**Figure 2 c** plots PD values i.a. extracted from **Fig. 2 b** following the applied colour coding. We derive a slope of -355 mV/nm for the PD values corresponding to the symmetric (t = b) barrier thickness increase (red to blue circles in **Fig. 2 c**), whereas the sole increase of b (t = 1.5 nm) yields a slope of -173 mV/nm (black triangles in **Fig. 2 b**). The inversion of the IFGARD stack (t ↔ b) does not significantly alter the PD value as indicated by the double triangle in **Fig. 2 c** (green and black) and the potential scans in **Fig. 2 b** that are coloured accordingly. **Figure 2 c** proves the fact that both, negative and positive PD values are accessible by the presented concept allowing the IFGARD to reach the desirable flat-band condition (compare **Fig. 1 f**) under any reasonable operating voltage in case of electrically driven devices.

Ultimately, the best barrier thickness constellations for the polarisation field guarding (t = 2.0 nm, b = 1.5 nm, or vice versa) boost the oscillator strength by a factor of 100 if compared to the conventional, 3-nm-high QD embedded in AlN. In other words, the photon rate provided by each of such GaN QDs is increased by two orders of magnitude. Nevertheless, the advantages of the IFGARD even go beyond such a tremendous increase in overall QD brilliance. The absence of the QCSE for the IFGARD case in **Fig. 1 d** leads to a QD emission energy of 4.2 eV, which is now exclusively governed by the confinement, whereas the conventional QD from **Fig. 1 a** emits at 3.5 eV due to the red-shift induced by the additional QCSE. In direct comparison to the 50%-higher QD from **Fig. 2** with emission energies of 2.9 eV and 4.0 eV, for the respective conventional and optimum IFGARD constellations, the QD size dependence of the emission energies is reduced by a factor of three from 3.5 eV - 2.9 eV = 0.6 eV to 4.2 eV - 4.0 eV = 0.2 eV. Hence, the energetically broad luminescence of conventional, e.g., nitride QD ensembles is minimized by the IFGARD. This is a fundamental prerequisite for any, e.g., laser application with a QD gain medium, as the QD dimensions will predominantly only affect their emission energy via the quantum confinement - a much less significant parameter if compared to the QCSE in nitrides.

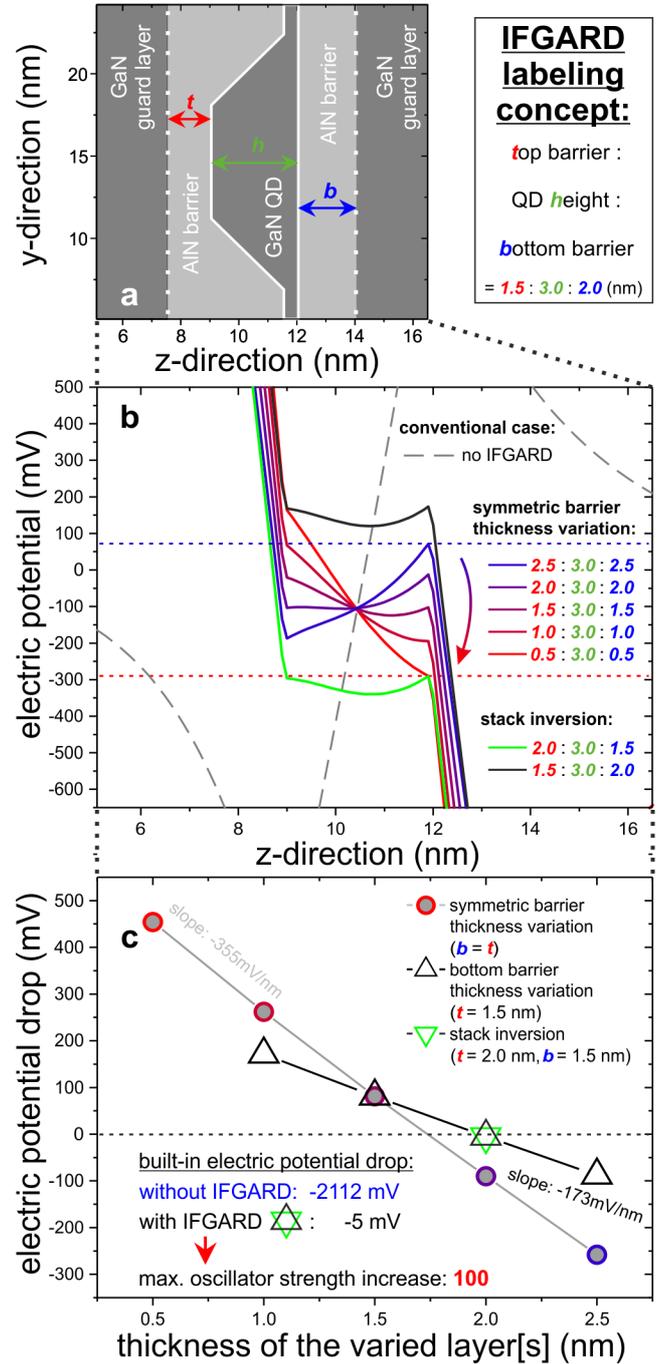

**Figure 2 | IFGARD tuning.** (a) Sketch of the IFGARD QD structure illustrating the AlN top (*t*) and bottom (*b*) barrier thickness along with the QD height (*h*). Piezo- and pyroelectric potential traces for *h* = 3 nm are shown in (b) for different AlN barrier thicknesses. Here, a symmetric barrier thickness variation (*t* = *b*) tunes the inclination of the potential gradients (blue → red), whereas a stack inversion with asymmetric barrier thicknesses (*t* ≠ *b*) only results in a potential offset, cf. (b). The entire electric potential drop inside of the particular QD is evaluated in (c) for several barrier thicknesses, indicating the optimum configuration by a double triangle. Vanishing of this electric potential drop causes an up to 100-times larger electron-hole oscillator strength, if compared to the conventional QD case.

## Field guarding in quantum wells

After having exemplified the basic field-guarding concept and even its tunability for the case of QDs, we now come to an intuitive explanation regarding the functionality of the IFGARD based on the QW structure exemplified in **Fig. 3 a**. Here in this figure, the GaN IFGARD QD from **Fig. 1 d** got replaced by a GaN QW, again exhibiting a horizontal orientation of the polar c-axis. Similar to the QD case in **Fig. 1 e**, interface-charges build up at each of the GaN/AlN or AlN/GaN interfaces of the IFGARD QW structure as illustrated in **Fig. 3 a** by the + (red) or - (black) signs. Due to this particular, reverse interface sequence of the IFGARD it is now feasible to achieve flat-band conditions inside of this single-QW as shown in **Fig. 3 b - top** (black line) for a 2-nm-thick, single GaN QW encapsulated by two AlN barrier layers each with a thickness of $t = b = 1$ nm. In comparison, the conventional QW ($h = 2$ nm) illustrated by the red, dotted line in **Fig. 3 b - top** exhibits a pronounced band-structure inclination.

Already the fundamental symmetry of this QW IFGARD structure brings an intuitive analogy into mind - a stack of open-circuit, plate-type capacitors as depicted in **Fig. 3 a**. In this analogy, the distance between the capacitor plates corresponds to the thicknesses of the GaN QW and the AlN barriers. The crystal's pyro- and piezoelectricity causes constant space charge densities at the interfaces of the IFGARD heterostructure similar to a charged, plate-type capacitor. Here, the central capacitor plates depicted in **Fig. 3 a** generate an electric field that can exactly be neutralized by the reversed field caused by the outer capacitor plates resulting in a field-free zone similar to the field-guarded interior of the IFGARD QW structure. Generally, the homogeneous electric field in between capacitor plates remains constant if their distance is varied and only the voltage ascribed to the potential difference in between the charged surfaces changes. In analogy, changing the thickness of the GaN QW or AlN barriers does not spoil the field-guarding effect as the relevant electric field superposition inside of the QW remains zero. Exactly the same observation is true for the electric field across a particular AlN barrier, which is in our analogy evoked by the charged, left *or* right plates of the inner *and* outer plate-type capacitor. Such a constant, non-zero electric field inside each of the AlN barriers is directly evidenced by a potential drop over a certain length interval in the corresponding band-structure calculations shown in **Fig. 3 b** and **c**. A constant slope of the band-structure trend (equivalent to constant electric fields) inside (non-zero) or outside (zero) of the AlN barriers is caused by the fixed space charges at all GaN/AlN and AlN/GaN interfaces evoked by pyro- and piezoelectricity. Interestingly, the analogy of stacked, open-circuit, plate-type capacitors facilitates a most simplistic understanding of QW IFGARD heterostructures as long as plan-parallel interfaces of infinite size are assumed.

The analogy gets into difficulties for the QD case as interfaces of different size occur in addition to less-polar QD side facets. Hence, any more complex nano-structure always requires the here applied numerical 3D-solution for the field situation as described in the **SI** in addition to all further simulation details. However, it is exactly the deviating interface geometry in the IFGARD QD structures that allows the tunability of the field inside of the QD by barrier thickness variations, cf. **Fig. 2**.

Generally, the fundamental IFGARD stack comprises exactly one barrier (e. g. AlN) along with one QW (e. g. GaN, see. **Fig. 3 a**) and can arbitrarily be repeated (counted by $n \in \mathbb{N}$) without sacrificing the beneficial IFGARD effect seen in, e.g., **Fig. 3 b - bottom** for a 1-nm-thick double-QW. Such IFGARD QW stacking is neither limited by the number of QWs, nor by the particular QW or barrier thickness as shown in **Fig. 3 c**. By repeating the IFGARD stack, one can create a step-wise multi QW structure exhibiting flat-band conditions within each QW. Here, the barrier thickness just regulates the height of each potential step. It is important to understand that any IFGARD stacking is not restricted to QW structures and can also be applied to any other type of nanostructure. Even more complex IFGARD band structure schemes become accessible as soon as the nanostructure composition is altered (e.g. within the $Al_xGa_{1-x}N$ system) in regard to the guard layers. As a result, highly unconventional and formerly inaccessible potential landscapes can be generated as depicted in **Fig. S2**.

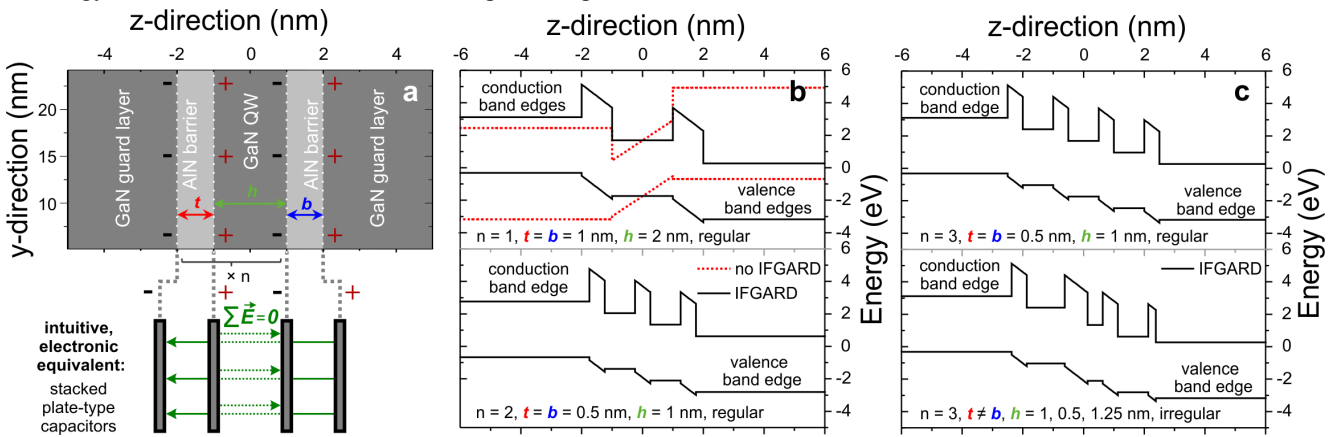

**Figure 3 | IFGARD QW.** The physics behind the field-guarding concept is explained based on its application to quantum wells (QWs) as illustrated in (a). Here, interface charges occur due to the built-in crystal polarisation (plus and minus signs), a situation similar to stacked, open-circuit plate-type capacitors, cf. (a). Again, the polar crystal axis is horizontal, matching the scan direction for the band edge profiles in (b) and (c). Due to the plane-parallel interfaces of infinite area, neither the QW (*h*) nor the barrier thickness (*t* and *b*) have an influence on the flat-band condition within the individual QW. Even multiple (n) stacks of the fundamental IFGARD layer sequence are achievable as exemplified for a regular double-, triple-, and an irregular triple-QW. As the flat-band conditions are always maintained, the barrier thickness exclusively governs the step-height in the band structure. Please see the **SI** for a variation of the QW composition.

### Discussion

Generally, the particular layer sequence of any IFGARD nanostructure can straightforwardly be realized based on well-established procedures as explained in the **SI**. As a result, numerous IFGARD advantages come into reach at low development expenses, clearly underlining the peculiarity of the entire concept. Here, the drastically increased device brightness is accompanied by a set of additional, pivotal benefits.

For the case of the laterally extended QWs, the electron-hole pair annihilation rate does not necessarily limit the quantum efficiency but still governs the ultimate operation speed rendering the light emission from polar QWs a rather slow phenomenon with limited light emission density. Any, e.g., electrical carrier injection into the matrix material surrounding the QW(s) leads to an electron-hole pair population that decays with a polarisation-limited rate, while the device's emission intensity relies on pump power in combination with a certain lateral size of the QW. Here, the main advantage of the IFGARD concept is the miniaturisation of lateral dimensions and the novel opportunity to further increase the pump power due to the enhanced device speed (no saturation), besides the aforementioned spectral narrowing of the emission. In contrast, as soon as smaller nanostructures like, e.g., zero-dimensional QDs are considered, a low electron-hole annihilation rate spoils the monochromatic emission as each QD must not be populated by more than one electron-hole pair. This, so-called, ground state exciton must decay with a rate that surpasses the QD fill rate. Otherwise, the formation of multi-excitons with deviating emission energies occurs.[18] In this context, the IFGARD-enhanced recombination rate of the ground state exciton indirectly suppresses parasitic channels, which boosts the quantum efficiency.

Therefore, charge carriers should remain in the matrix material for a timeframe governed by the electron-hole annihilation rate, which, however, enables a strong influence of parasitic decay channels. Hence, the IFGARD concept is of outmost value as soon as polar QDs[22-26] (e.g. [0001]-wurtzite GaN or [111]-zincblende InGaAs QDs) are considered, as both, the device speed and quantum efficiency can be raised.

Generally, the boost in electron-hole pair annihilation rate by the IFGARD originates from an improved electron and hole wavefunction overlap that also reduces the electric dipole moment.[27,28] As a direct consequence, the electrostatic coupling to charge fluctuations of nearby defects will be drastically reduced.[4,29] At the same time, the coupling to phonons diminishes[30,31] due to the reduced electron-hole separation, a most pivotal effect for electrically triggered, one- and two-photon sources[32,26] operating up to room-temperature. Hence, the emission of each individual QD will not only become brighter, but also more energetically defined, and less temperature-sensitive.[33] Such ultraviolet one- and two-photon sources represent an ideal candidate for the sub-diffraction analysis and nano-manipulation of extended molecules (DNA, RNA, etc.).[34] Interestingly, the absorbance of these molecules[35,36] nicely matches the emission range of GaN-based QDs,[37] enabling a strong perspective for individual bond cleavage[38,39] based on one- and two-photon absorption if combined with optical near-field techniques[40] in order to reach a spatial resolution that scales with the QD size only.

IFGARD LEDs and LDs directly raise the question for an electrical contact and bipolar doping. Here, it can be of great advantage that the outer guard-layers of the IFGARD-based structure comprise the same material as the nanostructure in the active region. Electrical contacts and the bipolar doping of, e.g., GaN are nowadays straightforwardly achievable,[2,41-43] whereas achievements of identical practicability are not yet accessible for AlN and cause excessive research efforts.[44-46] The electrical excitation of a single IFGARD QD is always based on a tunnelling process through the thin barrier layers comprising a material with a larger band gap. Therefore, the tunnelling probability is enhanced across the lateral extent of the QD and otherwise - in between the QDs - reduced due to the increased barrier thickness, cf. **Fig. 2 a**. In this sense, the IFGARD enables a current-channelling through the individual QDs, an effect that is otherwise achieved in single QD devices by complex processing of apertures.[47] Please note that exactly the same effect is also relevant for extended structures like, e. g., one-dimensional quantum wires.

From a fundamental point of view, the IFGARD can beneficially be applied to all semiconductor combinations, which exhibit strong piezo- and/or pyroelectric fields. As a result, the (optical) characteristics of such next generation structures based on the IFGARD will no longer be predominantly affected by the QCSE as most frequently reported for III-nitride nanostructures.[2,5,14-19,22,26-33] In general, the field-guarding conception boosts the radiative recombination and reduces the spectral emission width, while suppressing any parasitic recombination processes, an advantage that goes hand-in-hand with a strongly enhanced operation speed and a miniaturization of highly efficient (quantum) light sources.

**Funding acknowledgement** We acknowledge support from the Deutsche Forschungsgemeinschaft (DFG) within the Collaborative Research Center 787 and the European Union FP7-ICT project NEWLED, No. FP7- 318388.

**Author contributions** G. H., G. C., and A. H. are the inventors of the IFGARD. G. H. simulated all QD and S. W. all QW structures. G. C. and G. H. supervised the research endeavour and wrote the entire manuscript.

# Supplementary Information

## Shielding electrostatic fields in polar semiconductor nanostructures
G. M. O. Hönig, S. Westerkamp, A. Hoffmann, and G. Callsen

### Methodical Overview

It is the aim of this section to summarize our numerical procedure that facilitates the sophisticated, three-dimensional (3D) modelling for the particular case of quantum dots (QDs). Naturally, a principally equivalent, but one-dimensional (1D) method can trivially be applied to only two-dimensional structures like quantum wells (QWs). All calculations are based on an implementation of the 8-band-k·p formalism for wurtzite materials like nitrides, which is in detail described in literature and is further extended for the QD case by considering the particle interactions within the Hartree-Fock approximation.[1-3] In the following we will provide a brief explanation for the entire set of complex calculations for the case of a QD as shown in **Fig. S1**.

A common simulation starts with the creation of, e.g., the 3D QD structure embedded in the matrix material (**Fig. S1 a**), whose size, shape, and chemical composition are specified on a finite-differences grid. All these individual properties of the particular nanostructures were extracted from atomic force[4,5] and transmission electron microscopy[6] analyses enabling a truly realistic description of such nanostructures. For the QD calculations we chose a mesh width of 0.1 nm in a box-shaped computation area - a value well below the actual monolayer spacing in e.g. GaN[7] of ≈ 0.26 nm. A careful convergence behaviour analysis of all calculation results does not only confirm the chosen mesh width, but also the size of the calculation area of ≈ 30×30×20 nm³ (x×y×z) as adequate.[8] Please note that such large calculation areas are problematic for any atomistic calculations[9] but are essential in order to derive realistic (optical) properties in-line with various experimental results.[1,10,11]

The differing crystal lattice parameters that originate from the varying chemical composition in the calculation area create a strain distribution that is iteratively calculated using a continuum mechanical model[12,13] allowing a strain relaxation in the main growth direction known as the c-axis ([0001] direction). The resultant strain tensor distribution (e.g. $\varepsilon_{zz}$ along the [0001] direction is indicated in **Fig. S1 b**) affects the local electronic situation within and around the QD directly by strain-induced energy band deformations/shifts and indirectly by so-called piezoelectric polarisations. Subsequently, the calculation of the strain-dependent piezoelectric and the pyroelectric charge distributions inherent to the wurtzite structure[14] is performed, evoking a corresponding electrostatic potential (see **Fig. S1 c**) as described by the basic Poisson equation.

Generally, the careful treatment of such electrostatic potentials is also of outstanding importance for materials that crystalize in the zinc-blende configuration - a most relevant fact for instance for InGaAs/GaAs QDs. Recently, the [111] growth direction has been a matter of intense research efforts dedicated to InGaAs/GaAs.[15-17] Here, the application of the IFGARD concept could overcome the luminosity limitations of such [111]-based arsenide QDs, bringing their most prominent advantage for quantum photonics - a vanishing excitonic fine-structure splitting - to full bloom. Exactly in such arsenide materials, the second (quadratic) order of piezoelectricity has to be taken into account,[18-20] while the spontaneous pyroelectricity inherent to nitrides is absent.

As soon as the electrostatic potential is known for the entire calculation area, one can straightforwardly create the Hamiltonian matrix values for each segment of the mesh. Applying the local 8×8 Hamiltonian matrix[3,21] yields the coupling of the energetically lowest conduction band and the three topmost valence bands. Here, the previously derived electrostatic potential adds to the main diagonal of the 8×8 Hamiltonian. Furthermore, our Hamiltonian includes the effects of the spin-orbit and crystal-field coupling, which mainly affects the energy separations between the confined ground and excited hole states besides the band-mixing effects. A set of material parameters[22,23] is utilized for the entire calculations as further described by Winkelnkemper et al.,[24] whereas the sign of the piezoelectric constant $e_{15}$ has been corrected as explained by Tomić et al.[25] in reaction to the discussion in literature.[26,27] All relevant parameters are summarized in **Table S1**. Solving the Schrödinger equation yields single-particle electron and hole envelope wavefunctions[28] as depicted in **Fig. S1 d** - a first, but rather inadequate description of an electron-hole pair because the inherent interactions are still neglected.

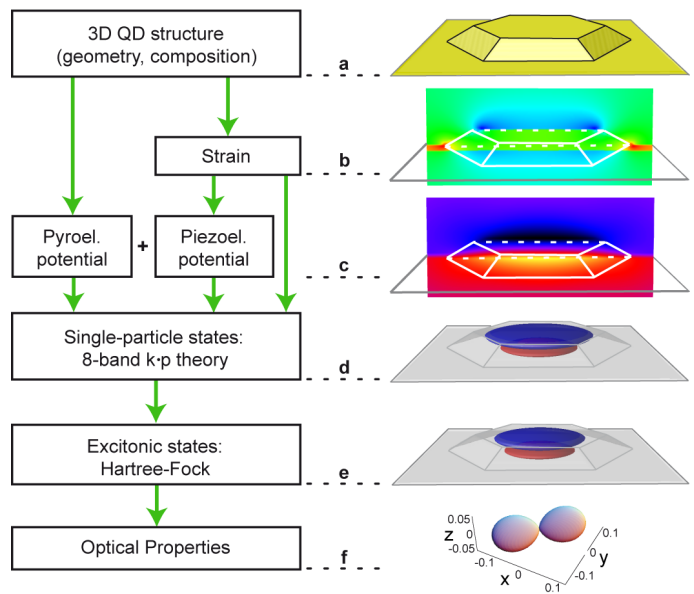

**Figure S1 | Simulations.** Calculation scheme for the applied 8-band k·p approach that yields the optical properties of the nanostructure of choice. (a) First, a nanostructure is defined before the strain distribution is continuum-mechanically calculated. The resulting $\varepsilon_{zz}$ strain tensor component along the [0001] direction is exemplarily indicated in (b). Subsequently, the electrostatic potential (c) is derived considering the effects of pyro- and piezoelectricity of the specific crystal lattice. Solving the 8×8 Hamiltonian including the coupling of the energetically lowest conduction and the three highest valence bands - 4×2 bands due to the spin projections - yields single particle electron and hole envelope wavefunctions (d). An additional consideration of particle interactions (Coulomb and non-classical exchange interaction) generates converged electron-hole densities (d). Finally, the optical properties of such a two-particle state approximation are determined, yielding e.g. the oscillator strength of the corresponding, optical transition (f).

| Param. | GaN | AlN | Param. | GaN | AlN | Param. | GaN | AlN | Param. | GaN | AlN |
|---|---|---|---|---|---|---|---|---|---|---|---|
| $a_{lc}$ (nm) | 0.3189 | 0.3112 | $e_{15}$ (Cm$^{-2}$) | -0.326 | -0.418 | $\Delta_{SB}$ (eV) | 0.017 | 0.019 | $a_2$ (eV) | -11.3 | -11.8 |
| $c_{lc}$ (nm) | 0.5185 | 0.4982 | $e_{31}$ (Cm$^{-2}$) | -0.527 | -0.536 | $m_{e\parallel}/m_0$ | 0.186 | 0.322 | $D_1$ (eV) | -3.7 | -17.1 |
| $E_{11}$ (GPa) | 390 | 396 | $e_{33}$ (Cm$^{-2}$) | 0.895 | 1.56 | $m_{e\perp}/m_0$ | 0.209 | 0.329 | $D_2$ (eV) | 4.5 | 7.9 |
| $E_{12}$ (GPa) | 145 | 137 | $P_{pyro}$ (Cm$^{-2}$) | -0.034 | -0.090 | $E_{P\parallel}$ (eV) | 17.292 | 16.927 | $D_3$ (eV) | 8.2 | 8.8 |
| $E_{13}$ (GPa) | 106 | 108 | $\varepsilon_r$ | 9.8 | 9.1 | $E_{P\perp}$ (eV) | 16.265 | 18.165 | $D_4$ (eV) | -4.1 | -3.9 |
| $E_{33}$ (GPa) | 398 | 373 | $E_g$ (eV) | 3.51 | 6.25 | $E_v$ (eV) | 0.8 | 0.0 | $D_5$ (eV) | -4.0 | -3.4 |
| $E_{44}$ (GPa) | 105 | 116 | $\Delta_{CF}$ (eV) | 0.010 | -0.169 | $a_1$ (eV) | -4.9 | -3.4 | $D_6$ (eV) | -5.5 | -3.4 |

**Table S1** | Compilation of material parameters for GaN and AlN used for the one- and three-dimensional 8-band-k·p simulations.

A more reasonable description of such a two-particle state known as an exciton is given by the mean field Hartree-Fock particle interaction approximation,[2] yielding bound electron-hole densities shown in **Fig. S1 e**. Here, the Coulomb- and the non-classical exchange interactions between the electron and hole are iteratively calculated leading to a renormalisation of both wavefunctions for each iteration step until the total excitonic energy is converged. Finally, the optical transition properties of the exciton(s) formed by the converged electron and hole wavefunctions are determined as exemplified in **Fig. S1 f**. Previously, we have successfully applied this entire numerical procedure to an in detail analysis of polar and non-polar QDs in perfect agreement with numerous experimental results.[6,29-32] Here, even an approximation of multi-excitons was recently derived[10] based on the Configuration Interaction (CI) method, which yields non-separable wavefunctions. Generally, QD exciton simulations require a 3D description, while QW band edge calculations can sufficiently be described within a standard 1D approach, allowing a doubling of the simulation resolution. In summary, all our simulations describe the optical properties for the nanostructure of choice based on well-established procedures as directly approved by the experimental evidence[6,29-32] - the supreme judge for modelling approaches.

## Stacking of field-guarded active regions

Generally, all modern LED and LD structures comprise extended layer stacks of various composition[33,34] in order to boost significant parameters as charge carrier injection, light out-coupling, quantum efficiency, etc. - all aiming towards a maximisation of the most pivotal device luminosity. Hence, it is a question of special importance to clarify whether the IFGARD is *generally* compatible with such extended layer stacks without treating all optimizations required for an entire device.

In order to illustrate this matter, **Fig. S2** shows the band edge profiles of two stacks of seven Al$_{1-x}$Ga$_x$N/AlN QWs (each AlN barrier has a thickness of 0.25 nm in order to approximate one monolayer). While in **Fig. S2 a** all QWs still consist of pure GaN, the gallium-content is arbitrarily reduced in some of the QWs of **Fig. S2 b** in order to demonstrate the general capability of the IFGARD concept. Here, the composition variation leads to a band edge tilt within all QWs, which is representative for the individual gallium/aluminium ratio. Most intriguingly, both, positive and negative band edge inclinations can be achieved, independent of the individual QW thickness. As a result, the IFGARD enables new pathways for the design of highly unconventional potential landscapes as hinted in **Fig. S2 b**. Suddenly, the band edge inclination inside of the nanostructure becomes a tunable parameter that can either be addressed by composition (QW case, see **Fig. S2 b**) or by the individual implementation of the IFGARD (QD case, see **Fig. 2**). Nevertheless, any particular band edge engineering for devices with an externally applied bias remains a task for future work and goes well beyond the scope of the present manuscript.

By realizing such larger numbers of stacked IFGARD QWs, nanowires, or QDs the different refractive indices of the active region, the barrier, and the guard materials can even be utilized for planar mode-guiding approaches not only in, e.g., 1D[35] and 2D[36] photonic crystals, but also in basic edge-emitting lasers in order to further improve the specific light emission characteristics of the device. Here, only a sufficient thickness of the entire IFGARD stack must be reached in order to achieve any mode-guiding towards the, e.g., device's side facets (perpendicular to the c-axis).

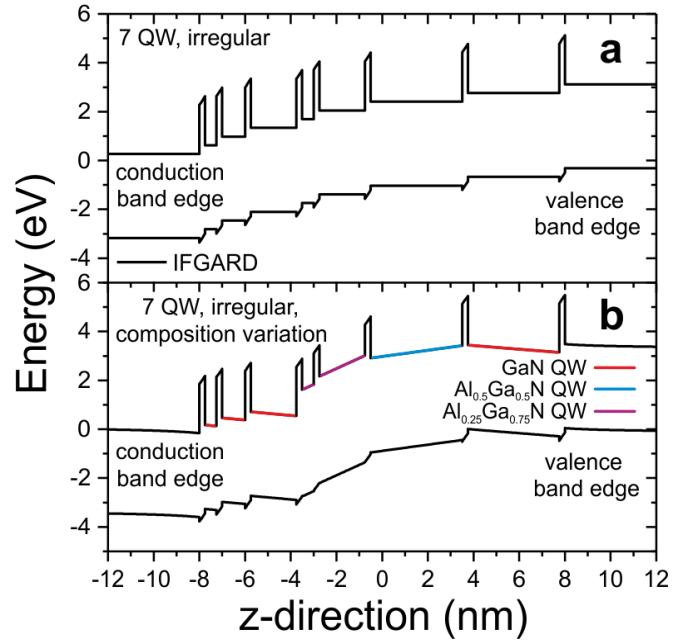

**Figure S2 | Composition tuning.** (a) The electric field guarding effect is maintained even for extended stacks of quantum wells (QWs) - here seven were chosen arbitrarily - as the flat-band conditions in the active region are still preserved. (b) Any composition variation in the QWs leads to a direct tunability of the immanent band structure inclination and step height. As a result, positive and negative band structure tilts become accessible independent of the individual QW thickness, only governed by the particular Al$_x$Ga$_{1-x}$N composition. For simplicity we chose an AlN barrier thickness of 0.25 nm to approximate the thinnest possible structure of one monolayer.

## Analysis of the technical feasibility

The growth of heterostructures that comprise QWs, quantum wires, or QDs along with numerous additional layers serving as Bragg reflectors, electron blocking layers, seed layers (polarity control), etc. is a well established procedure for many of the major, semiconductor compound families. Here, mainly the strongly polar oxide- and nitride-based correspondents often suffer from large, inherent electric fields as they most preferentially crystallize in the wurtzite structure. Directly related, highly sophisticated heterostructures have been developed throughout the last decades and comprise extended layer stacks featuring smooth interfaces.[37] Each of the layer thicknesses (guard layer, barrier, and QW) and, if applicable, QD geometries[4,38,39] that were assumed for our demonstration of the IFGARD in **Fig. 1 - 3** is straightforwardly achievable based on standard growth techniques as exemplified in the following section for the particular case of nitrides. Generally, such crystal polarisation is also highly relevant in (e.g. arsenide-based) cubic crystals. Here, the occurrence of piezoelectricity leads to similar detrimental effects if, e.g., the increasingly popular [111] growth direction is considered yielding major advantages for the generation of non-classical light based on QDs.[15-17,20]

Growing IFGARD heterostructures based on e.g. nitride material can directly be achieved as long as all layers do not exceed the critical thickness for plastic relaxation, characteristic for the specific material system. Here, as an example, **Fig. 2** assumes an AlN barrier thickness variation from 0.5 - 2.5 nm, well above the thickness of one monolayer and beneath the critical thickness of AlN grown on GaN of ≈ 3 nm.[40] Hence, as long as all layers are sufficiently thin, they are pseudomorphically strained and e.g. the growth of AlN on GaN is straightforwardly feasible, while the appearance of first cracks is reported for AlN thicknesses of 6 - 10 nm.[41-44] At the same time, the inclusion of, e.g., GaN QWs is well feasible regarding the corresponding thickness range[45-47] from **Fig. 3** and also the growth of matching guard layers is straightforward. Here, two possible main device categories are accessible for the IFGARD concept. First, the IFGARD stack can be grown on, e.g., a bulk GaN substrate[48] (bottom guard layer) and finally be capped by a sufficiently thin GaN layer (top guard layer) in order to ensure its optical transparency. Second, the entire IFGARD stack can be realized in a free-standing structure like, e.g., a nanobeam comprising a symmetric guard layer configuration with thicknesses scaling up to around 100 nm in order to achieve a sufficient mechanical stability of the final structure[35,49] along with a reasonable optical transparency (see the following section). We would like to remark that the particular thickness of the guard layers provides quite a flexible option for tailoring an IFGARD-based device as it is only the occurrence of the additional interfaces that ensures the entire functionality as depicted in **Fig. 3 a**.

Naturally, at first sight, the IFGARD only favours the inclusion of QWs as they are most preferentially, pseudomorphically strained, in contrast to QDs whose growth process itself often relies on strain relaxation. Hence, the common, so called, Stranski-Krastanov (SK) growth mode of nitride QDs[50,51] cannot straightforwardly be achievable in an IFGARD-based structure.[52] Here, the rather thin and pseudomorphically strained AlN barrier layers (see **Fig. 1 - 2**) do not provide a sufficient lattice miss-match for SK QDs. Nevertheless, most recent studies on, e.g., GaN QD growth have shown that the underlying growth mechanism of such QDs can strongly deviate from the SK mode. Here, a desorption-driven growth mode was reported[53] that does not rely anymore on common SK prerequisites, which is also true for the GaN QD nucleation close to structural defects.[54] Also common droplet epitaxy[55] can generate QD growth on pseudomorphically strained layers and is consequently well suited for any IFGARD QD device. Naturally, any electrical operation of the entire IFGARD structure is always based on charge carrier tunnelling through the sufficiently thin AlN barrier layers.

Tremendous efforts of the last years were dedicated to the site-controlled growth of e.g. GaN QDs[56] enabling novel quantum optical applications in the ultraviolet spectral range as described in the main article. Here, several techniques exist like, e.g., the QD nucleation in etch pits,[57] on nanowires,[58] or on strain apertures,[59] which all enable positioned, single QD(s) beyond the limits of a SK nucleation. In summary, all structural parameters for the active region (QWs and QDs), the barriers, and the guard layers that were assumed for a first demonstration of the IFGARD concept in **Fig. 1 - 3** are highly realistic and are not even limited to a particular nanostructure type or material system.

## Light absorption in the guard layers

At a first glance, the one and only apparent challenge of the IFGARD arises from the reabsorption of light in the guard layers as they comprise the same material as the active region in order to achieve flat-band conditions. Regarding the QD from **Fig. 2** we showed that the electron-hole pair annihilation rate is increased by a factor of 100. This advantage is now partially counterbalanced by the light reabsorption in the guard layers. For an example we assume 50-nm-thick guard layers, which, even in a totally freestanding structure as known from one- and two-dimensional photonic crystals, still results in an mechanically stable nano-device.[35,36] Hence, for typical GaN QD emission energies around 50% of the emitted light gets reabsorbed[60] in the top guard-layer - a tolerable effect if compared to the increase in electron-hole annihilation rate by 100 for the ideal IFGARD configuration, cf. **Fig. 2 b**. However, please note that this first, simplistic estimation does not account for the Fabry-Pérot interferences that occur in the guard layer(s). The optical transparency of the guard layers for the particular QD emission wavelength ($\lambda$) exhibits a modulation governed by the guard layer thickness, exhibiting optima at $\approx n \cdot \lambda/8$ ($n \in \mathbb{N}$).[61] However, such in detail design optimizations reside beyond the scope of this manuscript as also the characteristic radiation pattern of the QDs must be taken into account. Nevertheless, the trivial statement that any guard layer thickness of the order of the emitted wavelength enables sufficient transmission remains valid in consideration of the IFGARD benefits. Additionally, the light out-coupling along the c-axis can further be enhanced if the local density of optical states is altered based on a cavity structure as commonly applied for, e.g., nanobeam lasers.[35]